\documentclass[11pt,showpacs,preprintnumbers,amsmath,amssymb,prd,nofootinbib,superscriptaddress]{revtex4-2}

\usepackage{dcolumn}
\usepackage{bm}
\usepackage{ifpdf}
\usepackage{hyperref}
\usepackage{xcolor,color,graphicx,graphics,physics}
\usepackage[spanish,english]{babel}
\usepackage[latin1]{inputenc}
\usepackage[OT1]{fontenc}
\usepackage{latexsym,amssymb,amsmath,amsfonts, slashed,cancel}
\usepackage{makeidx}
\usepackage{epsfig,subfigure}
\usepackage{natbib}
\usepackage{epstopdf}   
\usepackage{mathrsfs}
\usepackage{hyperref}
\hypersetup{colorlinks=true, linkcolor=blue, citecolor=blue, urlcolor=blue}
\usepackage{enumerate}

\everymath{\displaystyle}
\usepackage{graphicx}

\usepackage[T1]{fontenc}
\usepackage{amsmath}
\usepackage{amssymb}
\usepackage{graphicx}
\usepackage{xcolor}

\newcommand{\bea}{\begin{eqnarray}}
\newcommand{\eea}{\end{eqnarray}}

\newcommand{\orcid}[1]{\href{https://orcid.org/#1}{\includegraphics[width=10pt]{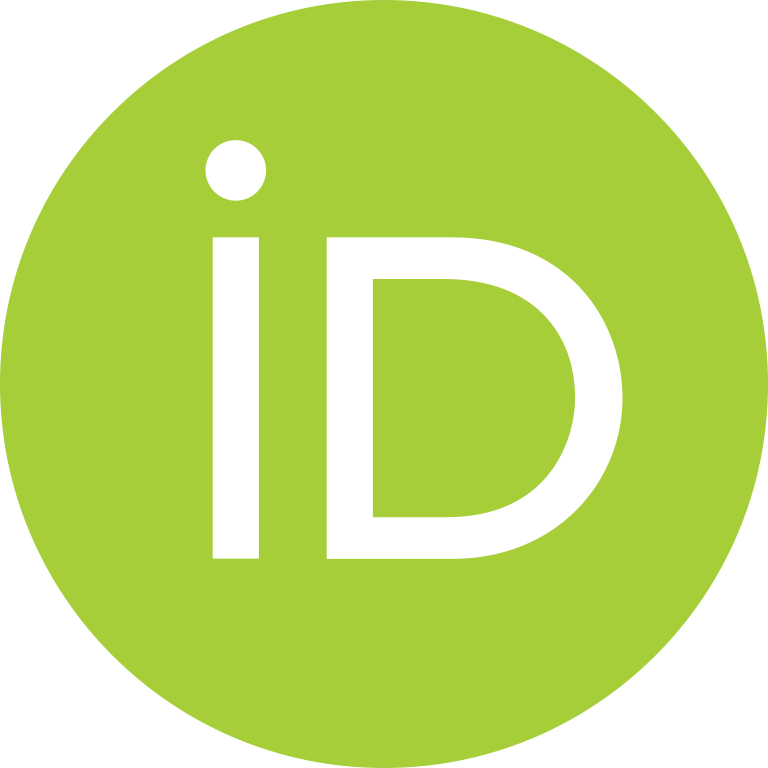}}}

\begin{document}

\title{Non-Hermitian electron-positron annihilation under thermal effects}

\author{D. S. Cabral  \orcid{0000-0002-7086-5582}}
\email{danielcabral@fisica.ufmt.br}
\affiliation{Programa de P\'{o}s-Gradua\c{c}\~{a}o em F\'{\i}sica, Instituto de F\'{\i}sica,\\ 
Universidade Federal de Mato Grosso, Cuiab\'{a}, Brasil}

\author{A. F. Santos \orcid{0000-0002-2505-5273}}
\email{alesandroferreira@fisica.ufmt.br}
\affiliation{Programa de P\'{o}s-Gradua\c{c}\~{a}o em F\'{\i}sica, Instituto de F\'{\i}sica,\\ 
Universidade Federal de Mato Grosso, Cuiab\'{a}, Brasil}

\author{R. Bufalo \orcid{0000-0003-1879-1560}}
\email{rodrigo.bufalo@ufla.br}
\affiliation{Departamento de F\'{\i}sica, Universidade Federal de Lavras,\\
37203-202, Lavras, Minas Gerais, Brazil}

\author{N. B. Xavier \orcid{0000-0000-0000-0000}}
\email{nathan.xavier1@estudante.ufla.br}
\affiliation{Departamento de F\'{\i}sica, Universidade Federal de Lavras,\\
37203-202, Lavras, Minas Gerais, Brazil}

\begin{abstract}
In this paper we examine the thermal effects into the $e^{+}e^{-}\to \ell^{+}\ell^{-}$ scattering in a non-hermitian extension of QED.
We compute the thermal contributions to this scattering cross-section within the Thermo Field Dynamics approach. 
In order to highlight the non-hermitian effects we have considered some limits of interest: i) zero-temperature limit and high-energy limit and ii) high-temperature regime.
Since this type of scattering possesses accurate experimental data for the cross-section (for muon and tau at the final state)  it can be used to set stringent bounds upon the non-hermitian parameters.

\end{abstract}

\maketitle

\section{Introduction}

In standard quantum theory, hermiticity is a fundamental property that ensures observables, such as energy, have real eigenvalues, corresponding to physically measurable quantities. This property not only guarantees real energy levels, preventing unphysical complex values, but also preserves the unitarity of the theory, which is essential for probability conservation. Howewer, some years ago, a new theory was developed in which hermiticity is no longer required to obtain real observables. Instead, this condition is replaced by spacetime reflection symmetry, or $\mathcal{PT}$\footnote{A combination of the discrete spacetime symmetries of parity $\mathcal{P}$ and time reversal $\mathcal{T}$ \cite{CP2}.}-symmetry, without sacrificing any of the essential physical principles of quantum mechanics. This new framework, known as non-Hermitian $\mathcal{PT}$-symmetric quantum theory, allows complex Hamiltonians to have spectra that are real and positive.

In recent years, non-Hermitian Hamiltonians with $\mathcal{PT}$-symmetry have garnered significant attention due to their intriguing and unexpected properties, such as the reality of the energy spectrum and unitarity \cite{Bender:1998ke,Mostafazadeh:2001jk,Bender:2005tb,Bender:2019,Alexandre:2023afi}. These characteristics have stimulated extensive research in this area of quantum theory \cite{Bender:2006fe,Makris_2008,longhi,Ashida:2017rso,El-Ganainy:2018ksn,Midya_2018,Matsumoto:2019are,Ashida:2020dkc}. The flexibility gained by relaxing Dirac's Hermitian condition, allowing new terms in both free and interaction components of a field theory under non-Hermiticity combined with $\mathcal{PT}$-symmetry, has proven phenomenologically attractive. This framework has led to numerous studies, including applications to neutrino mass and oscillation models in non-Hermitian QED and non-Hermitian Yukawa theory \cite{Alexandre:2015kra,Ohlsson:2015xsa,Alexandre:2017fpq,Ohlsson:2020idi,Alexandre:2020gah}, extensions of the Goldstone and Englert-Brout-Higgs mechanisms \cite{Mannheim}, and a $\mathcal{PT}$-symmetric extension of the Nambu-Jona-Lasinio (NJL) model in quantum chromodynamics \cite{NJL}. Additionally, there has been exploration into extending the CPT theorem to non-Hermitian Hamiltonians \cite{CPT}, examining Poincar\'{e} symmetries and representations in pseudo-Hermitian quantum field theory \cite{Poincare}, analyzing the effects of dissipative non-Hermitian terms on electrical transport in one-dimensional $\mathcal{PT}$-symmetric models with chiral symmetry \cite{Chiral}, and studying nonadiabatic transitions in 
$\mathcal{PT}$-symmetric two-level systems \cite{nonadiabatic}, among others. Although many properties of the non-Hermitian extension of QED have been investigated, its thermal behavior remains largely unexplored. In this work, we study the thermal effects on the $e^{+}e^{-} \to \ell^{+}\ell^{-}$ scattering process within a non-Hermitian extension of QED.

To incorporate temperature effects into quantum field theory, two main approaches are used: (i) the imaginary-time formalism, developed by Matsubara \cite{Matsubara}, and (ii) the real-time formalism, which has two variants: the closed-time path formalism \cite{Schwinger} and the Thermo Field Dynamics (TFD) \cite{Umezawa1, temp000, khannatfd}. In this work, we adopt the TFD formalism. This method involves doubling the original Fock space by introducing a dual space, or tilde space, and applying Bogoliubov transformations. The original and tilde spaces are linked through a mapping governed by the tilde conjugation rules. The Bogoliubov transformation acts as a rotation between these two spaces, embedding temperature effects. This approach results in a doubling of degrees of freedom, allowing Green's functions to be represented in a two-dimensional matrix structure. Notably, the real-time propagators are composed of two parts: one representing the zero-temperature propagator and the other, a temperature-dependent component.

This paper is organized as follows.
In Section \ref{sec2}, we examine the main aspects of the non-Hermitian QED model, we also discuss its gauge structure related with the issue of leptonic mass.
Section \ref{sec3} provides a brief overview of the TFD formalism.
In Section \ref{sec4}, we present the main definitions and aspects of the TFD formalism required to compute the transition amplitude and cross-section for the $e^{+}e^{-} \to \ell^{+}\ell^{-}$ scattering process at finite temperature.
In Section \ref{sec5}, we calculate the thermal cross-section at the leading order in terms of the non-hermitian parameters, we also analyze limits of physical interest.
Moreover, we use cross-section experimental data  for muon and tau final states to assess the axial coupling $\alpha_a$.
To highlight the non-Hermitian effects on the $e^{+}e^{-} \to \ell^{+}\ell^{-}$ cross-section, we compare the (zero temperature) non-Hermitian cross-section, the QED result, and the corresponding experimental data.
Finally, in Section \ref{secconclusion}, we present some concluding remarks. For the subsequent discussions and calculations, the following convention for the metric signature has been used: $\text{diag}{(\eta_{\mu\nu})}=(1,-1,-1,-1)$.

\section{Non-hermitian QED}
\label{sec2}

To evaluate scattering amplitudes in the TFD formalism, it is essential to obtain the solutions for the free fields as well as their completeness relations. Therefore, we begin by reviewing the main aspects of non-Hermitian QED \cite{Alexandre:2015kra,Alexandre:2017fpq}.

Our starting point is a non-hermitian extension of the free fermionic Lagrangian, given by
\begin{eqnarray}
\mathcal{L}=\overline{\psi}\left[\frac{1}{2}i\gamma^\alpha\overset{\leftrightarrow}{\partial}_\alpha-m-\mu\gamma^{5}\right]\psi,\label{eq01}
\end{eqnarray}
where the symmetric derivative is defined as $A\overset{\leftrightarrow}{\partial}_\alpha B=A\partial_\alpha B- (\partial_\alpha A)B$ and $\gamma^5$ denotes the usual axial matrix.
Actually, the presence of an axial mass term $\mu$ breaks the Lagrangian hermiticity and induces new and interesting features. 

Furthermore, the modified Dirac equation is obtained from the Lagrangian \eqref{eq01} and is given by
\begin{eqnarray}
\left(i\gamma^{\alpha}\partial_{\alpha}-m-\mu\gamma^{5}\right)\psi(x)=0.\label{eq02}
\end{eqnarray}
Hence, the dispersion relation for the fermionic field yields an effective mass
\begin{eqnarray} \label{disp_rel}
M= \sqrt{m^{2}-\mu^{2}}, \quad \mu \in \mathbb{R}.
\end{eqnarray}
Besides modifying the dispersion relation in terms of the mass $M$ \eqref{disp_rel}, the non-hermitian parameter $\mu$ controls the effective contribution from the left- and right-handed chiralities (see details in the section \ref{sec2.1}), so that the model   \eqref{eq01} has implications to neutrino physics \cite{Alexandre:2015kra,Alexandre:2017fpq}.
Moreover, the axial mass in \eqref{eq01} is responsible to stem phenomenological couplings, such as the electric dipole and toroidal moments, in the non-relativistic QED \cite{bufalo-2024}.

As usual, the solutions for the Dirac equation \eqref{eq02} can be expanded in plane waves, that is
\begin{align}
\psi\left(x\right)  = & \, \sum_{s}\int\frac{\text{d}^{3}p}{\left(2\pi\right)^{3}}\sqrt{\frac{M}{E_{p}}}\left[a_{p}^{s}u^{s}\left(p\right)e^{-ip\cdot x}+(b_{p}^{s})^\dagger\upsilon^{s}\left(p\right)e^{ip\cdot x}\right]\cr
 = &\,\psi^{+}(x)+\psi^{-}(x), \label{eq04a}\\
\overline{\psi}\left(x\right)  =& \, \sum_{s}\int\frac{\text{d}^{3}p}{\left(2\pi\right)^{3}}\sqrt{\frac{M}{E_{p}}}\left[b_{p}^{s}\overline{\upsilon}^{s}\left(p\right)e^{-ip\cdot x}+(a_{p}^{s})^{\dagger}\overline{u}^{s}\left(p\right)e^{ip\cdot x}\right]\cr 
 =& \, \overline{\psi}^{+}(x)+\overline{\psi}^{-}(x), \label{eq05a}
\end{align}
in which the upper index $(\pm)$ corresponds to the positive and negative energy modes, respectively.
Moreover, within the second quantization, we have that
\begin{align}
\left\{ a_{p}^{r},(a_{q}^{s})^{\dagger}\right\}  & =\left\{ b_{p}^{r},(b_{q}^{s})^{\dagger}\right\} =(2\pi)^{3}\delta^{3}(\vec{p}-\vec{q})\delta_{rs},\label{eq05}\\
\left\{ a_{p}^{s},(b_{q}^{s})^{\dagger}\right\}  & =\left\{ b_{p}^{s},(a_{q}^{s})^{\dagger}\right\} =0\label{eq06}
\end{align}
as the set of operators $(b^\dagger,b)$ and $(a^\dagger,a)$ are related with the fermion and anti-fermion, respectively.

From the solutions \eqref{eq04a} and \eqref{eq05a} we find the modified fermionic  completeness relations
\begin{align}
\mathcal{P}_u(p)=&\,\sum_{s}u^{s}(p)\overline{u}^{s}(p)  =  \frac{(m-\mu\gamma_{5})\slashed{p}+M^{2}}{2M^{2}}, \label{eq08P}\\
\mathcal{P}_v(p) =&\,\sum_{s}\upsilon^{s}(p)\overline{\upsilon}^{s}(p)  =  \frac{(m-\mu\gamma_{5})\slashed{p}-M^{2}}{2M^{2}}.\label{eq09P}
\end{align}
So that the free fermionic propagator is readily  obtained
\begin{equation} \label{eq10P}
iS\left(p\right)=i\frac{\slashed{p}+m-\mu\gamma^{5}}{p^{2}-M^{2}}.
\end{equation}

The easiest way to incorporate the electromagnetic coupling into the Lagrangian \eqref{eq01} is to examine its behavior under the local gauge transformations
\begin{eqnarray}
\begin{cases}
\psi\to\exp\left[i\left(g_{v}+g_{a}\gamma^{5}\right)\theta\right]\psi\\
\overline{\psi}\to\overline{\psi}\exp\left[i\left(-g_{v}+g_{a}\gamma^{5}\right)\theta\right]\
\end{cases}\label{eq12}
\end{eqnarray}
implying into
\begin{eqnarray}
\mathcal{L}\to\overline{\psi}\left[\frac{1}{2}i\gamma^\alpha\overset{\leftrightarrow}{\partial}_\alpha-\left(m+\mu\gamma^5\right)e^{i2g_a\gamma^5\theta}\right]\psi-\overline{\psi}\gamma^\alpha(\partial_\alpha\theta)\left(g_v+g_a\gamma^5\right)\psi\label{eq03}
\end{eqnarray}
in which the last term clearly breaks the gauge invariance.
Hence, we can define the non-hermitian QED in terms of
\begin{equation}
\mathcal{L}=\overline{\psi}\left[\frac{1}{2}i\gamma^\alpha\overset{\leftrightarrow}{\partial}_\alpha-m-\mu\gamma^{5}\right]\psi+\overline{\psi}\left[\gamma^\alpha A_\alpha\left(g_{v}+g_{a}\gamma^{5}\right)\right]\psi\label{eq04}
\end{equation}
which incorporates two electromagnetic couplings: a vector-axial (V-A) interaction with a coupling $g_a$ in addition to the electric one $g_v$.
Now, the full model \eqref{eq04} is invariant,  in the massless limit $m=\mu=0$, under the transformations \eqref{eq12} and $A_{\mu}\to A_{\mu}-\partial_{\mu}\theta$.

An interesting aspect of the interaction Lagrangian
\begin{eqnarray} \label{int}
\mathcal{L}_{\rm int}=\overline{\psi}\left[\slashed{A}\left(g_{v}+g_{a}\gamma^{5}\right)\right]\psi
\end{eqnarray}
is that it is hermitian, satisfying $\mathcal{L}_{\rm int}^\dagger=\mathcal{L}_{\rm int}$.
Thus, the non-hermiticity appears only in the free part of the non-hermitian QED, as shown in \eqref{eq04}.

Finally, the modified interaction vertex $\left\langle \overline{\psi}A\psi\right\rangle$ obtained from  \eqref{eq04} is cast as
\begin{equation}
\left\langle \overline{\psi}A_{\alpha}\psi\right\rangle =-i\gamma_{\alpha}\left(g_{v}+g_{a}\gamma^{5}\right),
\end{equation}
where the first term represents the usual vector coupling while the second one corresponds to the novel V-A coupling.

In a general context, an alternative term to the V-A coupling emerge from Lorentz violating (LV) QED models \cite{kosteleckycross, cabralnew, manoel} in terms of the coupling
\begin{eqnarray}
\mathcal{L}_d\sim d_{\alpha\beta}\overline{\psi}\gamma^\alpha A^\beta\gamma^5\psi.
\end{eqnarray}
This reproduces the V-A coupling in our case when the LV-tensor is the metric tensor, $d_{\alpha\beta}=\eta_{\alpha\beta}$. However, under this consideration, the similarity no longer signifies a violation of Lorentz symmetry. Moreover, the generation of electric dipole moment in QED can be achieved for the case of $d_{\alpha\beta}=\sigma_{\alpha\beta}$, in which $\sigma_{\alpha\beta} = \frac{i}{2} \left[\gamma_\alpha,\gamma_\beta\right]$ \cite{Gorghetto:2021luj}.

Furthermore, as discussed above, the gauge invariance of the model \eqref{eq04} is only manifest for the massless case ($m=\mu=0$).
Actually, this phenomenon can be understood in terms of the spontaneous symmetry breaking \cite{peskin}.
We will examine the subject in the next section.

\subsection{The problem of the leptonic mass}
\label{sec2.1}

In order to establish the framework for our discussion about the spontaneous symmetry breaking (SSB) in the non-hermitian model, let us consider the massless Dirac theory \eqref{eq01} and define the left- and right-handed spinors
\begin{eqnarray}
\psi_L=\frac{1-\gamma^5}{2}\psi,\quad\quad\psi_R=\frac{1+\gamma^5}{2}\psi,
\end{eqnarray}
where for brevity, we consider a single flavour fermion.
In this case, the model \eqref{eq01} takes the form
\begin{eqnarray}
\mathcal{L}_{\text{massless}}=\frac{1}{2}i\overline{\psi}_R\gamma^\alpha\overset{\leftrightarrow}{\partial}_\alpha\psi_R+\frac{1}{2}i\overline{\psi}_L\gamma^\alpha\overset{\leftrightarrow}{\partial}_\alpha\psi_L.\label{eq21}
\end{eqnarray}
Furthermore, introducing a modified Yukawa interaction \cite{Alexandre:2017fpq}, with a charged bosonic particle $\phi$, and its non-hermitian contribution, as
\begin{equation}
\mathcal{L}_{\text{Y}}=-G_v\left(\overline{\psi}_L\phi\psi_R+\overline{\psi}_R\phi^\dagger\psi_L\right)-G_a\left(\overline{\psi}_L\gamma^5\phi\psi_R+\overline{\psi}_R\gamma^5\phi^\dagger\psi_L\right),\label{eq19}
\end{equation}
in which $G_v$ and $G_a$ are the vector and axial coupling constants, respectively. The chiral term introduced into the model \eqref{eq19} is similar to those used to describe pseudo-vector particles, like a pion-nucleon scattering \cite{yukawachiral}. 

In addition, we consider that the dynamics of the boson (Higgs) field $\phi$ is given by the $\phi^4$ theory 
\begin{equation}
\mathcal{L}_\phi=\left({\partial}_\alpha\phi\right)^\dagger\left(\partial^\alpha\phi\right)-\frac{m_\phi^2}{2}\phi^\dagger\phi-\frac{f}{4}\left(\phi^\dagger\phi\right)^2.\label{eq20}
\end{equation}
Hence, the full model under consideration is written as
\begin{equation} \label{full_model}
\mathcal{L}_{\text{full}}=\mathcal{L}_{\text{massless}}+\mathcal{L}_{\text{Y}}+\mathcal{L}_\phi.
\end{equation}

On the other hand, by construction, when we consider the  $U(1)\otimes SU(2)$  symmetry, the transformation Eq.\eqref{eq12} is now cast as
\begin{equation}
U(1):\exp[-i\frac{(Y_v+Y_a\gamma^5)}{2}\theta_1];\quad\quad SU(2):\exp[-i\frac{(I_v+I_a\gamma^5)}{2}\vec{\sigma}\cdot\vec{\theta}_2],\label{eq16}
\end{equation}
for the transformation elements under these groups, respectively.

Moreover, on the relations \eqref{eq16}, we have that $Y_v$ and $Y_a$ are the hypercharges, while $I_v$ and $I_a$ are the magnitudes of the Weak Isospin $z-$component \cite{gellmann, nishijima}.
It is well-known that these quantities in the QED take the values: $(I_v,Y_v)=(-1/2,-1)$ to the left-handed and $(I_v,Y_v)=(0,-2)$ to the right-handed  fermion, and also $(I_v,Y_v)=(-1/2,-1)$ to the Higgs boson \cite{peskin, ryder}.
Actually, these quantities are related in terms of the Gell-Mann--Nishijima relation \cite{gellmannnishijima}
\begin{eqnarray}
Q_v=I_v+\frac{Y_v}{2}
\end{eqnarray}
where $Q_v$ is the electric charge.

It is important to note that an analogous relation can be established for the axial components of the transformations \eqref{eq16}, where we define $Q_a$ as the chiral charge in such a way that the fields $\psi_L$, $\psi_R$ and $\phi$ assume as values $-1$, $1$ and $0$, respectively.

Therefore, in other words, we can write the following values: $(I_a,Y_a)=(0,-2)$ to the left-handed and $(I_a,Y_a)=(0,2)$ to the right-handed fermion, and also $(I_a,Y_a)=(0,0)$ to the Higgs boson.
Taking these results into consideration, we obtain that the fundamental fields transform explicitly as
\begin{equation}
\psi_L\to \exp[i\frac{(1+2\gamma^5)}{2}\theta_1]\psi_L;\quad\quad\psi_R\to \exp[i(1-\gamma^5)\theta_1]\psi_R;\quad\quad\phi\to\exp[\frac{i}{2}\theta_1]\phi\label{eq17}
\end{equation}
under the $U(1)$ symmetry group, and
\begin{equation}
\psi_L\to \exp[-\frac{i}{4}\vec{\sigma}\cdot\vec{\theta}_2]\psi_L;\quad\quad\psi_R\to \psi_R;\quad\quad\phi\to\exp[-\frac{i}{4}\vec{\sigma}\cdot\vec{\theta}_2]\phi\label{eq18}
\end{equation}
under the $SU(2)$ group.
Therefore, strictly speaking, the set of transformations \eqref{eq17} and \eqref{eq18} show that the fermionic fields transform with an additional chiral contribution.

Actually, the Yukawa and Higgs Lagrangians \eqref{eq19} and \eqref{eq20} are gauge invariant under these transformations.
Hence, in order to make the full model \eqref{full_model} invariant, the Dirac massless theory \eqref{eq21} must be gauged in terms of the full covariant derivative 
\begin{eqnarray}
\nabla_\alpha=\partial_\alpha-i(a+b\gamma^5)X_\alpha-i\frac{c}{2}\vec{\sigma}\cdot\vec{Z}_\alpha
\end{eqnarray}
where  the coupling constants $a$, $b$ and $c$, in addition to the gauge fields $X_\alpha$ and $\vec{Z}_\alpha$, completely characterize the theory.

Finally, we can now apply all this setup to the non-hermitian case defined in terms of the Lagrangian \eqref{eq04}.
In order to implement the spontaneous symmetry breaking, we can choose the point $\phi\to\phi+v$, where $v$ is a zero of the effective Klein-Gordon potential \eqref{eq20},  that is $V_{\rm eff}(v)=0$ (actually,  $v=\langle\phi\rangle$ is the vacuum expectation value of the Higgs field).
Under these circumstances, the Yukawa couplings \eqref{eq19} are modified to
\begin{eqnarray} \label{eq19a}
\mathcal{L}_{\text{Y}}\to\mathcal{L}_{\text{Y}}-G_vv\left(\overline{\psi}_L\psi_R+\overline{\psi}_R\psi_L\right)-G_av\left(\overline{\psi}_L\gamma^5\psi_R+\overline{\psi}_R\gamma^5\psi_L\right).
\end{eqnarray}
At last, by comparing the full SSB model \eqref{full_model}, written in terms of \eqref{eq19a}, with the non-hermitian theory \eqref{eq04} we can observe that the fermionic masses are identified as
\begin{eqnarray}
m=G_vv,\quad \text{and}\quad  \mu=G_av.
\end{eqnarray}
This result shows that the non-hermitian Dirac theory is originally gauge invariant and massless.
Then, the scalar boson $\phi$ gives mass to the modified QED \eqref{eq04} from a non-hermitian Higgs mechanism, defined in a (single flavour) electroweak-like theory.

Having discussed and established the main properties of the non-Hermitian QED model in \eqref{eq04}, we now turn to our primary focus: examining the thermal properties of electron-positron scattering within this non-Hermitian framework. We introduce temperature using the TFD formalism, which allows for the analysis and calculation of thermal amplitudes in a manner similar to zero-temperature field theory.

\section{Thermo Field dynamics}
\label{sec3}

The main idea behind the TFD formalism is that we can interpret thermal vacuum expectation values of an arbitrary operator as its statistical average \cite{temp000, khannatfd}.
Formally, we can write
        \begin{align}
                \expval{\cal A}{0(\beta)}=Z^{-1}(\beta)\sum_ne^{-\beta E_n}\expval{\cal A}{n}.
        \end{align}
This analysis can be carried out through a thermal ground state $\ket{0(\beta)}$.
The definition of this state is only possible due to the construction of a thermal Hilbert space ${\cal H}_T$ in terms of the direct product of the original Hilbert space ${\cal H}$ and the doubled (tilde) Hilbert space $\tilde{\cal H}$, that is, ${\cal H}_T = {\cal H} \otimes \tilde{\cal H}$ \cite{temp000, khannatfd}.
        
Moreover, this formalism has two essential elements: (i) doubling of variables, essentially due to the introduction of $\tilde{\cal H}$ (physically speaking, the second Fock space is interpreted as a heat bath that ensures the system to stay in equilibrium), and (ii) the use of the Bogoliubov transformation, necessary to introduce thermal effects.

A direct consequence of the first observation is that we have the tilde conjugation rules which, for an arbitrary operator ${\cal A}$, are written as
        \begin{align}
                \widetilde{\left({\cal A}_i{\cal A}_j\right)} &=\tilde{\cal A}_i\tilde{\cal A}_j,\quad
               \widetilde{ \left(c{\cal A}_i+{\cal A}_j\right)}=c^*\widetilde{\cal A}_i+\widetilde{\cal A}_j, \\
               \widetilde{ ({\cal A}_i^\dagger)}&=\tilde{\cal A}_i^\dagger, \quad
               \widetilde{ (\widetilde{\cal A}_i)}=-\xi{\cal A}_i,\quad \left[{\cal A}_i,\widetilde{\cal A}_j\right]=0,\label{conj_rule}
        \end{align}
in which $\xi = \pm 1$ for fermions and bosons, respectively.
The thermal ground state is formally defined as follows
        \begin{align}
                \ket{0(\beta)}=\mathcal{U}(\beta)\ket{0,\widetilde{0}},
        \end{align}
where $\mathcal{U}(\beta)$ is the Bogoliubov operator of the thermal transformation, responsible to introduce the temperature, and defined as
        \begin{align}
                \mathcal{U}(\beta)=e^{-iG(\beta)},
        \end{align}
in which $G(\beta)=i\theta(\beta)\left(\widetilde{a}a-a^\dagger\widetilde{a}^\dagger\right)$.
In this way, considering fermionic fields, the thermal annihilation and creation operators (analogous relations hold for $b(p, \beta)$) are cast as
        \begin{eqnarray}
                a\left(p,\beta\right) &=e^{-iG(\beta)}a(p)e^{iG(\beta)},\quad 
                a^\dagger\left(p,\beta\right)=e^{-iG(\beta)}a^\dagger(p)e^{iG(\beta)}\label{6},
        \end{eqnarray}
        with similar expressions for the tilde operators.
Alternatively, we can expand the operator $U(\beta)$, so that we can rewrite \eqref{6} as the following
        \begin{eqnarray}
                a(p,\beta)&=U(\beta)a(p)-V(\beta)\tilde{a}^\dagger(p), \quad 
                a^\dagger(p,\beta)&=U(\beta)a^\dagger(p)-V(\beta)\tilde{a}(p)\label{8},
        \end{eqnarray}
where we have introduced the functions $(U,V)$ in terms of the Fermi-Dirac distribution $n_F(E)$, which explicitly read
        \begin{equation}
                U(\beta)=\left(1+e^{-\beta E}\right)^{-1/2} = \sqrt{e^{\beta E}n_F(E)}, \quad V(\beta)=\left(1+e^{\beta E}\right)^{-1/2} = \sqrt{n_F(E)}.
        \end{equation}
It is worth mentioning that, to obtain these relations for the tilde operators, it is sufficient to apply the tilde conjugation rules \eqref{conj_rule} to the expressions \eqref{8}.

Moreover, the anti-commutation relations for the thermal operators are written as
        \begin{align}
                \{a_p^r(\beta),(a_{q}^s(\beta))^\dagger\}=(2\pi)^3\delta^3(\vec{p}-\vec{q})\delta_{rs},
        \end{align}
have the same form as those for $T=0$ given in Eqs. \eqref{eq05} and \eqref{eq06}.

Since we want to compute finite temperature effects into a fermionic scattering, we can consider the above development and formally define the thermal amplitude as
\begin{equation} \label{amplitude}
\mathcal{M}\left(\beta\right)=\left\langle f,\beta\left|\hat{S}\right|i,\beta\right\rangle
\end{equation}
where the $\hat{S}$-matrix is given by
\begin{equation} \label{smatrix}
\hat{S}=\sum_{n=0}^{\infty}\frac{\left(-i\right)^{n}}{n!}\int d^{4}x_{1}d^{4}x_{2}\ldots d^{4}x_{n}\,\mathcal{T}\left[\hat{\mathcal{L}}_{\textrm{int}}\left(x_{1}\right)\hat{\mathcal{L}}_{\textrm{int}}\left(x_{2}\right)\ldots\hat{\mathcal{L}}_{\textrm{int}}\left(x_{n}\right)\right]
\end{equation}
in which $\mathcal{T} $ is the time ordering operator and
\begin{equation} \label{tilde_lag}
\hat{\mathcal{L}}_{\textrm{int}}\left(x_{n}\right)=\mathcal{L}_{\textrm{int}}\left(x_{n}\right)-\widetilde{\mathcal{L}}_{\textrm{int}}\left(x_{n}\right)
\end{equation}
describes the Lagrangian interaction part in the doubled notation of the TFD formalism.
Finally, we can define the differential cross section,   at the centre of mass frame, as usual
\begin{eqnarray} \label{diff_cross}
\frac{d\sigma}{d\Omega}=\frac{|\vec{p}_i||\vec{p}_f|}{(4\pi s)^2}\langle|\mathcal{M}|^2\rangle,
\end{eqnarray}
where $\langle|\mathcal{M}|^2\rangle$  represents the squared modulus of the transition amplitude \eqref{amplitude}, and the Mandelstam variable $s$ denotes the square of the center-of-mass energy (invariant mass). In addition, $\vec{p}_i$ and $\vec{p}_f$ are the 3-momenta of the initial and final particles, respectively.

\section{Thermal scattering process}
\label{sec4}

\begin{figure}[t!]
\centering
\includegraphics[scale=0.3]{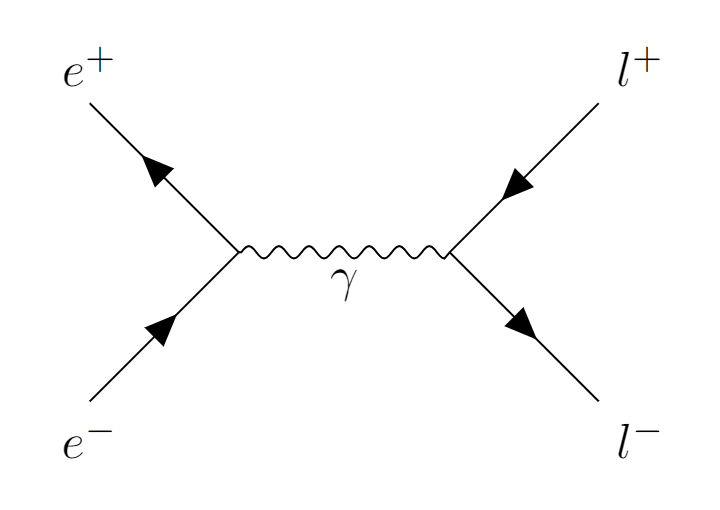}
\caption{Feynman diagram of the scattering at tree level.}
\label{fig1}
\end{figure}

Now that we have established the main definitions and aspects of the TFD formalism we are able to calculate the transition amplitude and the cross-section for the $e^{+}e^{-}\to \ell^{+}\ell^{-}$ scattering at finite temperature.
The tree-level Feynman diagram contributing to this process is depicted in Figure \ref{fig1}.
The first point to evaluate the amplitude \eqref{amplitude} is to define the corresponding initial and final states for this process as
\begin{eqnarray}
\ket{i(\beta)}&=&\sqrt{2E_{p_i}}\sqrt{2E_{k_i}}(a^{s_1}_{p_i}(\beta))^\dagger (b^{s_2}_{k_i}(\beta))^\dagger\ket{0(\beta)},\\\ket{f(\beta)}&=&\sqrt{2E_{p_f}}\sqrt{2E_{k_f}}(a^{s_3}_{p_f}(\beta))^\dagger(b^{s_4}_{k_f}(\beta))^\dagger\ket{0(\beta)},
\end{eqnarray}
where the factor $\sqrt{2E}$ is for normalization purposes.

Moreover, we can consider the scattering matrix \eqref{smatrix} and its  second-order element 
\begin{eqnarray}
\hat{S}^{(2)}=\frac{(-i)^2}{2!}\int d^4xd^4y \mathcal{T}\left\{\hat{\mathcal{L}}_I(x)\hat{\mathcal{L}}_I(y)\right\}
\end{eqnarray}
which allows us to compute the amplitude \eqref{amplitude} explicitly
\begin{align}
\mathcal{M}&=-\frac{1}{2}\int d^4xd^4y\bra{f(\beta)}\mathcal{T}\left\{\overline{\psi}_x\left[\slashed{A}_x\left(g_{v}+g_{a}\gamma^{5}\right)\right]\psi_x\overline{\psi}_y\left[\slashed{A}_y\left(g_{v}+g_{a}\gamma^{5}\right)\right]\psi_y\right\}\ket{i(\beta)}\nonumber\\
&=-\int d^4xd^4y\bra{f(\beta)}:\overline{\psi}_x^{+}\left[\gamma_\mu\left(g_{v}+g_{a}\gamma^{5}\right)\right]\psi_x^{+}\overline{\psi}_y^{-}\left[\gamma_\nu\left(g_{v}+g_{a}\gamma^{5}\right)\right]\psi_y^{-}:\ket{i(\beta)}\Delta_F^{\mu\nu}(x,y;\beta) \label{amplitude_2}
\end{align} 
where $:\ldots:$ stands for the usual Wick's normal ordering, while $\psi^{(\pm)}$ and $\overline{\psi}^{(\pm)}$ are the fermionic positive and negative energy solutions given in eqs \eqref{eq04a} and \eqref{eq05a}. Moreover, the gauge field propagator is defined as
\begin{eqnarray}
\Delta_F^{\mu\nu}(x,y;\beta) =-i\int \frac{d^4q}{(2\pi)^4}e^{-iq(x-y)}D^{\mu\nu}(q;\beta) 
\end{eqnarray}
in which the finite temperature propagator $D^{\mu\nu}(q;\beta) =\eta^{\mu\nu}\Delta_\beta (q)$ have the following expression \cite{khannatfd}
\begin{eqnarray}
\Delta_\beta (q) =\frac{1}{q^2}-\frac{2\pi i}{e^{\beta|q_0|}-1}\delta(q^2)
\end{eqnarray}
one can observe that the thermal effects are encoded in the second term.

Hence, evaluating the expectation value in the amplitude \eqref{amplitude_2}, also considering the above remarks, we are able to obtain
\begin{eqnarray}
\mathcal{M}=-4M_e M_{\ell} F(\beta)\overline{v}^{s_2}(k_1)\Gamma_\mu u^{s_1}(p_i) \,\Delta_\beta (s)\, \overline{u}^{s_3}(p_f)\Gamma^\mu v^{s_4}(k_f)
\end{eqnarray}
in which we have defined $\Gamma_\mu=\left[\gamma_\mu(g_v+g_a\gamma^5)\right]$, so that $\left(\overline{u}_1\Gamma_\mu u_2\right)^\dagger=\overline{u}_2\Gamma_\mu u_1$. 
In addition, all thermal distributions are contained in the factor
\begin{eqnarray}
F(\beta)=U_{p_i}U_{p_f}U_{k_i}U_{k_f}.
\end{eqnarray}

Finally, taking these results into account, we are able to compute the squared modulus of the
transition amplitude
\begin{eqnarray}
\langle|\mathcal{M}|^2\rangle = 4M_e^2 M_{\ell}^2 F^2(\beta)|\Delta_\beta(s)|^2\Tr{\mathcal{P}_u(p_i)\Gamma^\mu \mathcal{P}_v(k_i)\Gamma^\nu}\Tr{\mathcal{P}_v(k_f)\Gamma_\mu \mathcal{P}_u(p_f)\Gamma_\nu},\label{eq23}
\end{eqnarray}
 where the definitions \eqref{eq08P} and \eqref{eq09P} have been used.

Since the resulting integration over the phase space variables, to compute the cross section \eqref{diff_cross} for this scattering, results in an extensive function of the effective masses $M_e$ and $M_\ell$, it is convenient to consider some approximations upon the expression \eqref{eq23} in order to select the leading contributions.

Hence, we can observe the presence of the corrected mass $M$ \eqref{disp_rel} in the expression \eqref{eq23} and look to the non-expansion-based parameter $\mu$ as a small perturbation which enables us to establish bounds on the non-Hermitian effects.
So that we can use the parametrization $\mu_{\ell}=\lambda_{\ell} m_{\ell} $, with $\lambda_{\ell}$ being a small parameter $\lambda_{\ell} \ll 1$, which yields
\begin{eqnarray}
M_{\ell}^2 = m_{\ell}^2(1-\lambda_{\ell}^2).\label{eq09}
\end{eqnarray}
Actually, looking carefully to Eq. (\ref{eq09}), some considerations are in order about the axial leptonic mass contributions to the scattering amplitude \eqref{eq23}: we expect to obtain a clear protocol to separate the leading contributions of the masses when heavy leptons are considered in the scattering.

\subsection{Mass corrections}
\label{sec4.1}

In order to establish some approximations to evaluate the cross-section (for the amplitude \eqref{eq23}) consistently, we can consider the leading contribution of the mass correction parameter $\lambda_{\ell}$ into the electron modified mass \eqref{eq09} so that
\begin{eqnarray}
M_e=m_e\left(1-\frac{\lambda_e^2}{2}\right).
\end{eqnarray}
Moreover, since the relative standard uncertainty of the value of the electron's mass is of $u_r^{(e)}=3\times 10^{-10}$ \cite{masscorrections}, we have that
\begin{eqnarray}
\frac{m_e\lambda_e^2}{2}\leq 3\times10^{-10}\;m_e \longrightarrow\lambda_e\leq 2.45\times10^{-5}.
\end{eqnarray}
In an analogous way, we can obtain similar bounds also to the heavier leptons, for the muon we find
\begin{eqnarray}
\lambda_\mu\leq2.1\times 10^{-4},
\end{eqnarray}
in which its uncertainty is $u_r^{(\mu)}=2.2\times10^{-8}$ \cite{masscorrections}, while for the tau we obtain
\begin{eqnarray}
\lambda_\tau\leq1.2\times 10^{-2},
\end{eqnarray}
where the mass uncertainty reads $u_r^{(\tau)}=6.8\times10^{-5}$ \cite{masscorrections}.

We summarize these results of the mass corrections $\mu_{\ell}=\lambda_{\ell} m_{\ell}$ in the table \ref{tab1}.
In comparison, one can see that, effectively, the electron mass and its correction are negligible when compared with other heavier leptons (muon and tau).
Hence, we shall take $M_e=m_e \to 0$ in the next calculations.
\begin{table}[ht]
\begin{tabular}{|c|c|c|c|}\hline
Particle & Measured mass $(m_\ell)$ & Mass correction $(\mu_\ell)$ &  Correction parameter $(\lambda_\ell)$ \\\hline
Electron & $0.511$ MeV & $1.25\times10^{-5}$ MeV &  $2.45\times10^{-5}$ \\\hline
Muon & $105.658$ MeV & $2.22\times10^{-2}$ MeV &  $2.1\times10^{-4}$ \\\hline
Tau & $1776.86$ MeV & $21.32$ MeV & $1.2\times10^{-2}$ \\\hline
\end{tabular}
\caption{Relation of the physical mass $m_{\ell}$ and its axial correction $\mu_{\ell}$ for all leptons.}
\label{tab1}
\end{table}

\section{Thermal cross section}
\label{sec5}

Since we have established a proper procedure to select the leading mass contributions on the (squared) amplitude  \eqref{eq23}, we are now able to compute the differential cross section \eqref{diff_cross}.
Moreover, in order to evaluate the integration of the cross-section we will consider the center-of-mass frame, in which the Mandelstam variables are cast 
\begin{eqnarray}
s=E_{\rm cm}^2\qquad\text{and}\quad\quad t=m_\ell^2-\frac{\sqrt{s}}{2}\left(\sqrt{s}-2p_f\cos\theta\right).
\end{eqnarray}  
Hence, under these considerations, we obtain
\begin{eqnarray}
\left(\frac{d\sigma}{d\Omega}\right)_{\beta}&=&{F^2(\beta)}\left[\frac{1}{s^2}+\frac{(2\pi s)^2\delta^2(s)}{(e^{\beta \sqrt{s}}-1)^2}\right]\left(\frac{d\sigma}{d\Omega}\right). \label{diff_sec}
\end{eqnarray}
This expression showcases the strength of the TFD formalism, because up to the moment we have used standard canonical methods to compute the transition amplitude, and naturally the thermal contributions were separated as a factor.
The thermal factor in this frame of reference is
\begin{eqnarray}
F(\beta)=\frac{1}{4}\biggl[1+\tanh{\biggl(\frac{\beta \sqrt{s}}{2}\biggr)}\biggr]^2.
\end{eqnarray}
Furthermore, the non-thermal part of \eqref{diff_sec} is given by
\begin{align}
\left(\frac{d\sigma}{d\Omega}\right)=&\,\frac{g_a^4\varepsilon}{128\pi^2s}\left[s\left(2\lambda_\ell^2+1\right)\Xi_{11}(2\theta)-2m_{\ell}^2\left(2\Xi_{11}(2\theta)+\lambda_\ell^2\Xi_{12}(2\theta)\right)\right]\nonumber\\
&+\frac{(g_a^3g_v+g_ag_v^3)\varepsilon}{32\pi^2s}\lambda_\ell\left[4\Xi_{00}(\theta)\left(2m_\ell^2\Xi_{00}(\theta)-\varepsilon\right)-s\Xi_{11}(2\theta)\right]\nonumber\\
&+\frac{g_a^2g_v^2}{64\pi^2\varepsilon}\Big[16m_\ell^4\left(\lambda_\ell^2+2\right)\Xi_{00}^2(\theta)+(2\lambda_\ell^2+1)s\left(8\varepsilon\Xi_{00}(\theta)+s\Xi_{11}(2\theta)\right) \nonumber\\
& -2m_\ell^2\lambda_\ell^2\left(16\varepsilon\Xi_{00}(\theta)+5s\Xi_{11}(2\theta)-4s\right)+16\varepsilon\Xi_{00}(\theta)+4s\left(\Xi_{11}(2\theta)-1\right)\Big]\nonumber\\
&+\frac{g_v^4}{128\pi^2\varepsilon}\Big[8m_\ell^4\left(\lambda_\ell\Xi_{13}(2\theta)+4\left(\Xi_{00}^2(\theta)-1\right)\right)-2sm_\ell^2\left(8\Xi_{00}^2(\theta) +5\lambda_\ell^2\Xi_{11}(2\theta)\right)\nonumber\\
& +s^2(\lambda_\ell^2+1)\Xi_{13}(2\theta)\Big]
\end{align}
in which we can observe that the modified electron mass $m_e$ is absent due to our previous discussion (i.e. $m_e \ll m_\ell$ and $\lambda_e \ll \lambda _\ell$). Here, we have defined, by simplicity of notation, the angular dependence in terms of the quantity
\begin{eqnarray}
\Xi_{ab}(\theta)=\cos(\theta)+a\left(2b+1\right),
\end{eqnarray}
and the squared energy quantity
\begin{eqnarray}
\varepsilon=\sqrt{s\left(s-4m_\ell^2\right)}.
\end{eqnarray}
Finally, the remaining angular integration over the solid angle $d\Omega$ in the expresion \eqref{diff_sec} can be readily  performed, so that  the cross-section for this scattering reads
\begin{eqnarray}
\sigma_{\beta}(s)&=&{F^2(\beta)}\frac{\chi(s)}{s^2}\label{eq07}
\end{eqnarray}
in which the factor $\chi(s)$ has the following expression
\begin{align}
\chi(s)=&\,\frac{g_a^4\varepsilon}{12\pi s}\left[4m_\ell^2\left(\lambda_\ell^2-1\right)+s\left(\lambda_\ell^2+1\right)\right]-\frac{g_a^3g_v+g_v^3g_a}{3\pi s}\lambda_\ell\left[\varepsilon(s-m_\ell^2)\right]\nonumber\\
&+\frac{g_a^2g_v^2}{6\pi\varepsilon}\left[2m_\ell^4(\lambda_\ell^2+2)-sm_\ell^2(7\lambda_\ell^2+5)+s^2(2\lambda_\ell^2+1)\right]\nonumber\\&+\frac{g_v^4}{12\pi\varepsilon}\left[4m_\ell^4\left(5\lambda_\ell^2-2\right)-2sm_\ell^2\left(5\lambda_\ell^2+1\right)+s^2\left(2\lambda_\ell^2+1\right)\right]. \label{eq08}
\end{align}

It is important to note that the result obtained here from Eq.\eqref{eq08}, is general and incorporates QED corrections by introducing temperature effects and non-Hermiticity through the axial-vector coupling constant $g_a$ and the chiral parameter correction $\lambda_\ell$ (up to second order). To validate this framework, it is essential to analyze all relevant limits, including those associated with temperature and non-Hermitian modifications.

Some remarks about the cross section \eqref{eq07}  are in order: In one hand, in the zero-temperature limit ($\beta\to\infty$) we have that $F(\beta)=1$ and the expression \eqref{eq07}  becomes
\begin{eqnarray}
\sigma_{\beta\to\infty}(s)=\frac{\chi(s)}{s^2}.\label{eq13}
\end{eqnarray}

In addition, within the  high energy limit ($\sqrt{s} \gg m_\ell$, which effectively yields $m_{\ell}\to 0$), the non-thermal vertex correction will given exclusively by
\begin{eqnarray}
\sigma_{\rm HE}=\lim_{m_{\ell}\to0}\frac{\chi(s)}{s^2}=\frac{1}{s} \Delta (\alpha,\alpha_a,\lambda_\ell) \label{eq10}
\end{eqnarray}
in which we have defined
\begin{equation}
\Delta (\alpha,\alpha_a,\lambda_\ell) = \frac{4\pi}{3}\left(\alpha+\alpha_a\right)\left[(\alpha+\alpha_a)-4\sqrt{\alpha\alpha_a}\lambda_\ell+2(\alpha+\alpha_a)\lambda_\ell^2\right],
\end{equation}
where $\alpha=g_v^2/4\pi$ and $\alpha_a=g_a^2/4\pi$ represents the usual QED and axial fine structure constants, respectively.
One can also observe that the first term corresponds to the usual QED contribution \cite{peskin, ryder}, while the remaining are corrections due to the additional axial mass and V-A coupling.

On the other hand, at the high-temperature regime (in which $\beta \sqrt{s} \ll 1$) we have the following leading thermal behavior
\begin{equation}\label{high_temp}
\sigma_{\rm high-T} \simeq \frac{1}{8} \Delta (\alpha,\alpha_a,\lambda_\ell) \left[ \frac{\beta }{\sqrt{s}} + \frac{3 \beta^2 }{4 } +\cdots \right],
\end{equation}
where temperature independent terms were discarded.
We see a significant departure (in terms of its dependence on the squared energy $s$) of the high-temperature expression \eqref{high_temp} when compared with the $T=0$ result \eqref{eq10}.

Notice that we have establish bounds on the thermal expressions of the cross-section, allowing us to estimate the chiral parameters at a given temperature.

To accurately determine these limits, a comparison with experimental data is required. Moreover, since the cross-section for the scattering process $e^{+}e^{-}\to \ell^{+}\ell^{-}$ is known with high precision for final states involving muon and tau particles-but without accounting for temperature effects-we use the zero-temperature expression \eqref{eq10}, the experimental cross-section data, and the values of $\lambda_\ell$ from Table \ref{tab1} to place stringent bounds on the V-A coupling $g_a$.

\subsection{Muon Lepton}

Since we wish to assess the non-hermitian corrections to the $e^{+}e^{-}\to \mu^{+}\mu^{-}$  cross-section, by matching these contributions  with the associated error for the respective cross-section, we should recast the expression \eqref{eq10}  in the form
\begin{equation} \label{eq11}
\delta\sigma^{{\rm non-H}}\equiv \frac{ |\sigma_{{\rm MEAS}}-\sigma_{{\rm QED}}|}{\sigma_{{\rm QED}}}= \frac{1}{\alpha^2}\left|\alpha_a\left(2\alpha+\alpha_a\right)-4\sqrt{\alpha\alpha_a}(\alpha+\alpha_a)\lambda_\ell+2(\alpha+\alpha_a)^2\lambda_\ell^2\right|,
\end{equation}
in which the QED cross-section is given by $\sigma_{\rm QED}=4\pi\alpha^2/3s$, and also  $\sigma_{\text{MEAS}}$ is the experimental (or measured) value of the respective cross-section for $e^{+}e^{-}\to \mu^{+}\mu^{-}$ scattering,  obtained from different experiments according to Table \ref{tab2}.
In summary, we compare the non-hermitian deviation $\delta\sigma^{{\rm non-H}}$ with the experimental error $\delta\sigma^{{\rm MEAS}}$, so that we can set a bound upon the axial coupling $g_a$. 
We present a series of bounds upon $\alpha_a$, for different values of energy $\sqrt{s}$, in the Table \ref{tab2}. 
One can consider the average of the results and obtain that $\alpha_a^{(\mu)}=4.98\times10^{-4}$, which implies $\alpha_a^{(\mu)}\approx1/2008$, which is significantly smaller than the fine structure constant $\alpha_a^{(\mu)} \ll \alpha \simeq 1/137$ and is consistent with our idea to consider the non-hermitian effects as perturbations.

\begin{table}[ht]
\begin{tabular}{|c|c|c|c|c|c|c|c|} \hline $\sqrt{s}$ (GeV) & $\sigma_{\text{MEAS}}$ \newline (pb) & $\alpha_a^{(\mu)}$ ($10^{-4}$) & Reference & $\sqrt{s}$ (GeV) & $\sigma_{\text{MEAS}}$ \newline (pb) & $\alpha_a^{(\mu)}$ ($10^{-4}$) & Reference \\ \hline 13.9 & 472.7 $\pm$ 36.0 & 1.8 & \cite{muon35} & 56.0 & 30.9 $\pm$ 3.5 & 4.1 & \cite{muontau52} \\ \hline 22.3 & 184.7 $\pm$ 15.7 & 2.1 & \cite{muon35} & 56.5 & 19.9 $\pm$ 7.1 & 9.2 & \cite{muontauadditional} \\ \hline 34.5 & 73.2 $\pm$ 1.5 & 0.1 & \cite{muon35} & 57.0 & 21.7 $\pm$ 3.8 & 6.6 & \cite{muontauadditional} \\ \hline 35.0 & 66.1 $\pm$ 1.3 & 2.5 & \cite{muon35} & 57.8 & 27.5 $\pm$ 0.6 & 2.1 & \cite{muontau57} \\ \hline 38.3 & 56.4 $\pm$ 4.4 & 1.7 & \cite{muon35} & 58.0 & 25.1 $\pm$ 1.0 & 1.0 & \cite{muontauadditional} \\ \hline 43.6 & 42.0 $\pm$ 1.7 & 2.9 & \cite{muon35} & 58.7 & 32.2 $\pm$ 5.1 & 9.5 & \cite{muontauadditional} \\ \hline 52.0 & 33.5 $\pm$ 4.7 & 1.5 & \cite{muontau52} & 60.0 & 28.3 $\pm$ 4.6 & 6.1 & \cite{muontauadditional} \\ \hline 54.0 & 18.0 $\pm$ 8.0 & 13 & \cite{muontauadditional} & 60.8 & 28.3 $\pm$ 4.2 & 7.1 & \cite{muontauadditional} \\ \hline 55.0 & 23.5 $\pm$ 3.9 & 6.4 & \cite{muontau52} & 61.4 & 14.5 $\pm$ 3.2 & 12 & \cite{muontauadditional} \\ \hline \end{tabular}
\caption{Non-hermitian contributions to the $e^{+}e^{-}\to \mu^{+}\mu^{-}$  cross-section. We present the values of the axial coupling $\alpha_a$ in terms of the error $\delta\sigma^{{\rm MEAS}}$ obtained from eq.~\eqref{eq11}.}
\label{tab2}
\end{table}

\begin{figure}[ht]
 \includegraphics[scale=0.65]{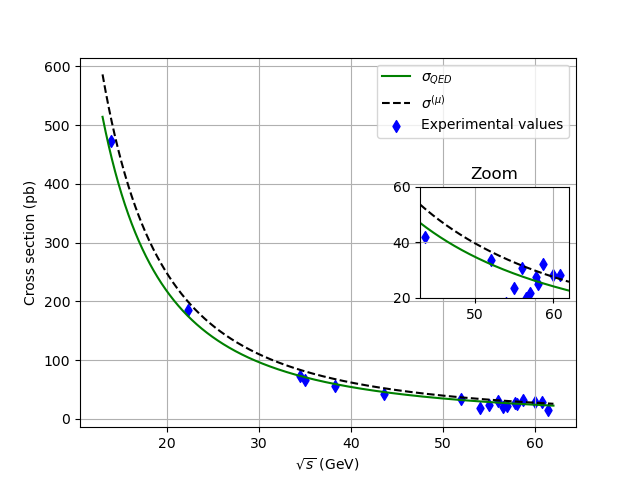}
 \caption{A comparison among the (zero temperature) non-hermitian cross section (dashed line) \eqref{eq13}, the QED result (green solid line) for the $e^{+}e^{-}\to \mu^{+}\mu^{-}$ scattering and the respective experimental data (table \ref{tab2}).}
 \label{fig2}
 \end{figure}

Moreover, in order to highlight the non-hermitian effects on the $e^{+}e^{-}\to \mu^{+}\mu^{-}$  cross-section, we present a plot of the complete cross section \eqref{eq13} and the usual QED value $\sigma_{\rm QED}$ in comparison with the experimental data (table \ref{tab2}), these are depicted in Figure \ref{fig2}.  In this plot was used $g_a=0.079$ with $\lambda_\mu=2.1\times10^{-4}$. We observe that the QED and the non-hermitian curves differ at low and intermediate energies ($\sqrt{s} < 50 \,{\rm GeV}$), with a better description of the data by the QED in the $30 \,{\rm GeV}\leq \sqrt{s} \leq 50 \,{\rm GeV}$ region; while  in the $\sqrt{s} > 50 \,{\rm GeV}$ region both curves have good agreement with the data.

\subsection{Tau lepton}

As previously discussed in section~\ref{sec4.1}, regarding the mass correction, the non-Hermitian contributions arising from the final states of muons and taus exhibit significant differences.

We now extend the previous analysis of the axial coupling $\alpha_a$, based on \eqref{eq11}, to the process $e^{+}e^{-}\to \tau^{+}\tau^{-}$.
A series of bounds on $\alpha_a$ for different energy values, $\sqrt{s}$, is presented in Table \ref{tab3}. 
By averaging these values, we obtain $\alpha_a^{(\tau)}=4.02\times10^{-4}$, which corresponds to $\alpha_a^{(\tau)}\approx1/2487$. This result is notably smaller than the fine-structure constant, satisfying $\alpha_a^{(\mu)} \ll \alpha \simeq 1/137$, and aligns with our approach of treating non-Hermitian effects as perturbative corrections.

\begin{figure}[ht]
 \includegraphics[scale=0.65]{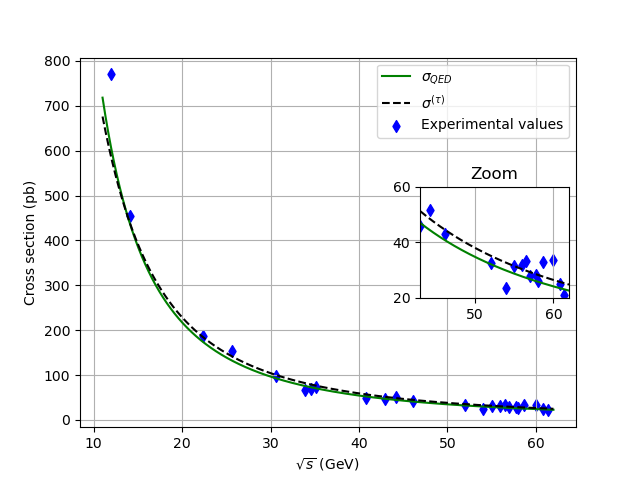}
 \caption{A comparison among the (zero temperature) non-hermitian cross section (dashed line) \eqref{eq13}, the QED result (green solid line) for the $e^{+}e^{-}\to \tau^{+}\tau^{-}$ scattering and the respective experimental data (table \ref{tab3}).}
 \label{fig3}
 \end{figure}

The non-Hermitian contributions to the $e^{+}e^{-}\to \tau^{+}\tau^{-}$ cross-section are illustrated through a comparison between the complete cross-section given by \eqref{eq13} and the standard QED prediction, $\sigma_{\rm QED}$, against the experimental data from Table \ref{tab3}. These results are depicted in Figure \ref{fig3}.  In this plot, the values $g_a=0.071$ and $\lambda_\tau=1.2\times10^{-2}$ were used.
\begin{table}[ht]
\begin{tabular}{|c|c|c|c|c|c|c|c|} \hline $\sqrt{s}$ (GeV) & $\sigma_{\text{MEAS}}$ (pb)& $\alpha_a ^{(\tau)}(10^{-4})$ & Reference & $\sqrt{s}$ (GeV) & $\sigma_{\text{MEAS}}$ (pb) & $\alpha_a^{(\tau)} (10^{-4})$ & Reference \\ \hline 12 & 770 $\pm$ 144 & 9.5 & \cite{taujade} & 52 & 32.5 $\pm$ 5.6 & 0.4 & \cite{muontauadditional} \\ \hline 14.1 & 454.3 $\pm$ 56.8 & 1.4 & \cite{taumarkj} & 54 & 23.6 $\pm$ 11.8 & 7.3 & \cite{muontauadditional} \\ \hline 22.4 & 186.9 $\pm$ 19.0 & 2.8 & \cite{taumarkj} & 55 & 31.5 $\pm$ 5.6 & 3.4 & \cite{muontau52} \\ \hline 25.6 & 154 $\pm$ 21.4 & 5.7 & \cite{taujade} & 56 & 31.8 $\pm$ 4.3 & 5.2 & \cite{muontau52} \\ \hline 30.6 & 98.2 $\pm$ 9.1 & 2.1 & \cite{taujade} & 56.5 & 33.2 $\pm$ 11.7 & 7.6 & \cite{muontauadditional} \\ \hline 33.9 & 66.5 $\pm$ 6.0 & 4.3 & \cite{taumarkj} & 57 & 27.9 $\pm$ 5.5 & 1.6 & \cite{muontauadditional} \\ \hline 34.6 & 69.7 $\pm$ 1.4 & 1.4 & \cite{taujade} & 57.8 & 28.3 $\pm$ 0.87 & 3.1 & \cite{muontau57} \\ \hline 35.1 & 72.6 $\pm$ 5.6 & 1.1 & \cite{taumarkj} & 58.0 & 25.9 $\pm$ 1.3 & 0.1 & \cite{muontauadditional} \\ \hline 40.8 & 48.5 $\pm$ 4.7 & 2.6 & \cite{taumarkj} & 58.7 & 32.8 $\pm$ 6.6 & 10 & \cite{muontauadditional} \\ \hline 43 & 45.9 $\pm$ 1.7 & 0.9 & \cite{taujade} & 60.0 & 33.5 $\pm$ 6.3 & 13 & \cite{muontauadditional} \\ \hline 44.2 & 51.6 $\pm$ 4.4 & 5.9 & \cite{taumarkj} & 60.8 & 25.1 $\pm$ 5.0 & 2.4 & \cite{muontauadditional} \\ \hline 46.1 & 42.9 $\pm$ 6.5 & 1.8 & \cite{taumarkj} & 61.4 & 21.2 $\pm$ 5.0 & 2.9 & \cite{muontauadditional} \\ \hline \end{tabular}
\caption{Non-hermitian contributions to the $e^{+}e^{-}\to \tau^{+}\tau^{-}$  cross-section. We present the values of the axial coupling $\alpha_a$ in terms of the error $\delta\sigma^{{\rm MEAS}}$ obtained from eq.~\eqref{eq11}.}
\label{tab3}
\end{table}

It is observed that the QED and non-Hermitian curves differ primarily at intermediate energy ranges, specifically for $ 20 \,{\rm GeV}< \sqrt{s} < 40 \,{\rm GeV}$, where the non-Hermitian model provides a better fit to the data in the $20 \,{\rm GeV}\leq \sqrt{s} \leq 30 \,{\rm GeV}$ region. However, for $\sqrt{s} > 40 \,{\rm GeV}$, both models exhibit good agreement with the experimental data.

\section{Conclusions}
\label{secconclusion}

In this work we have considered thermal effects into the $e^+ e^- \to \ell^+ \ell^-$ scattering described by a non-hermitian extension of the QED within the Thermo Field Dynamics approach.
We started by presenting a modified Dirac equation in terms of a non-hermitian axial mass, in which we have determined their plane waves solutions as well as the modified fermionic completeness relations.
As usual, the couplings were introduced in terms of the gauge invariance, which is only manifest in the massless limit.
Actually, since this phenomenon can be understood in terms of the spontaneous symmetry breaking, we revised the Higgs mechanism applied to this non-hermitian model.
It is worth to mention that in this non-hermitian QED, the free fermionic propagator and also the vertex function were both modified by an axial mass and a V-A coupling, respectively.

Since the TFD formalism allows to compute transition amplitudes exactly as in the $T = 0$ field theory, by doubling the set of operators and its respective Fock space (where the
second set acts like a heat bath), we have explicitly evaluated the scattering matrix element related
with the process $e^+ e^- \to \ell^+ \ell^-$, showing in details how the temperature effects are incorporated.
Nonetheless, the cross section for this scattering gives an extensive expression,  hence it was necessary to consider approximations on the expression \eqref{eq23} in order to select the leading contributions.
We implemented this idea by treating the non-expansion-based parameter $\mu$ as a small perturbation, which allows us to obtain bounds on the non-Hermiticity through the introduced parameter $\lambda_\ell$.
So that we can use the parametrization $\mu_{\ell}=\lambda_{\ell} m_{\ell} $, with $\lambda_{\ell}$ being a small parameter $\lambda_{\ell} \ll 1$; which allowed us to obtain bounds upon the non-hermitian parameter $\lambda_{\ell} $ and also led to a (flavor) hierarchy $\lambda_{e} \ll \lambda_{\mu} \ll \lambda_{\tau}$ on the leading contributions, allowing us to consider $M_e  \to 0$ on the calculations.

After computing explicitly the cross-section (for an arbitrary flavor final state) we have considered some  limits of interest in order to highlight its physical behavior: i) since the experimental data for the $e^+ e^- \to \ell^+ \ell^-$ cross section  (at $T=0$)  is known to great accuracy, we have applied the zero-temperature limit and high-energy limit, since such a comparison is only feasible in the zero-temperature regime; moreover, ii) we also evaluated the high-temperature regime of the cross-section, which shows a significant departure (in terms of the energy $\sqrt{s}$) when compared with the $T=0$ result. 
Actually, we used the experimental data for the cross-section in the $T=0$ limit to establish stringent bounds upon the V-A coupling $g_a$: we found that for muon $\alpha_a^{(\mu)}\approx1/2008$, while for tau $\alpha_a^{(\tau)}\approx1/2487$.
Moreover, we presented a plot to highlight the non-hermitian and QED theories in terms of the experimental data: we observed that the QED and the non-hermitian curves have good agreement with the data at high-energies, and that both curves differ (considerably) at intermediate energies (for the muon $30 \,{\rm GeV}\leq \sqrt{s} \leq 50 \,{\rm GeV}$ and for the tau $20 \,{\rm GeV}\leq \sqrt{s} \leq 30 \,{\rm GeV}$ region). Although some, but not all, key results are expressed in terms of zero-temperature quantities, the most significant findings and discussions presented here are framed within a thermal model. This approach provides all the necessary tools and mechanisms to derive the corresponding temperature-dependent versions of these results.

Based on the findings of this analysis, we intend to extend our investigation on thermal aspects of non-hermitian QFT to other scattering processes at tree level, as well as to explore loop corrections in future studies \cite{future}.

\section*{Acknowledgments}

This work by A. F. S. is partially supported by National Council for Scientific and Technological
Development - CNPq project No. 312406/2023-1. D. S. C. thanks CAPES for financial support.
R.B. acknowledges partial support from Conselho
Nacional de Desenvolvimento Cient\'ifico e Tecnol\'ogico (CNPq Project No.~ 306769/2022-0).

\section*{Data Availability Statement}

No Data associated in the manuscript.

\section*{Conflicts of Interest}

No conflict of interests in this paper.


\global\long\def\link#1#2{\href{http://eudml.org/#1}{#2}}
 \global\long\def\doi#1#2{\href{http://dx.doi.org/#1}{#2}}
 \global\long\def\arXiv#1#2{\href{http://arxiv.org/abs/#1}{arXiv:#1 [#2]}}
 \global\long\def\arXivOld#1{\href{http://arxiv.org/abs/#1}{arXiv:#1}}


\end{document}